\documentstyle[amssymb,aps,prb,epsf]{revtex}
\begin{document}

\twocolumn[\hsize\textwidth\columnwidth\hsize\csname
@twocolumnfalse\endcsname

\title{Phase behavior of a system of particles with core collapse}
\author{E. A. Jagla\cite{correo}}
\address{Centro At\'omico Bariloche\\
Comisi\'on Nacional de Energ\'{\i}a At\'omica\\
8400 S. C. de Bariloche, R\'{\i}o Negro, Argentina}
\maketitle

\begin{abstract}
The pressure-temperature phase diagram of a one-component system, with
particles interacting through a spherically symmetric pair potential in 
two dimensions is
studied. The interaction consists of a hard core plus an additional
repulsion at low energies. It is shown that at zero temperature, instead of
the expected isostructural transition due to core collapse occurring when
increasing pressure, the system passes through a series of ground states
that are not triangular lattices. In particular, and depending on parameters,
structures with squares, chains, hexagons and even quasicrystalline ground 
states are found.
At finite temperatures the solid-fluid coexistence line presents a zone with
negative slope (which implies melting with decreasing in volume) and the fluid
phase has a temperature of maximum density, similar to that in water.
\end{abstract}

\vskip2pc] \narrowtext

\section{Introduction}

Determination of the phase structure of real materials from first principles
calculations has been one of the aims of statistical mechanics since long
ago. Although a qualitative understanding of the processes leading to the
different kinds of phase transitions (between gas, liquid, and one or more
solid phases) in the pressure-temperature ($P$-$T$) phase diagram of a
classical system has been gained, it is clear that the quantitative fitting
of the behavior of real materials requires a detailed knowledge of the
interaction between particles and a great deal of computational work, that
only in recent years has become feasible.

In addition to the usual materials in which atoms or molecules are the basic
constituents, in the last years colloidal dispersions have provided a new
kind of systems in which parameters such as particle size and interaction
potential can be varied greatly.\cite{coloides} These systems
consist of a set of latex spheres in colloidal suspension, with the
aggregate of some amount of non-adsorbing polymer, which modifies the
interaction potential between the particles. Their study has practical importance 
in relation to the properties of many common 
substances (such as ink, paints, cosmetics, blood, etc.). It is clear that a
knowledge of the phase behavior of different model systems is important in order 
to compare the theoretical predictions with the experimental results.

Much effort has been spent in the elucidation of the properties of binary
mixtures of particles of two different sizes, where segregation,
flocculation, partial crystallization and other phenomena may occur.\cite
{binary} On the other hand, other studies have been directed towards the
determination of the phase behavior of identical particles interacting
through different model potentials. In this case the possibilities for the
behavior of the system are not so wide as in the case of binary mixtures
but, however, interesting phenomena occur. It was shown for instance, that
the usual solid-liquid-gas phase diagram of particles interacting through a
hard core repulsion plus a long range attraction is modified when the range
of the attraction is decreased.\cite{pocito} More precisely, the liquid-gas
coexistence curve disappears if the range of the attractive potential is
lower than about 30\% of the hard core radius. More interestingly, when the
range of the attractive potential is reduced below about 8\% of the
repulsive range, a coexistence curve separating two isostructural solid
phases appear. 

A more obvious isostructural transition occurs in the case in which the
attractive well is replaced by a repulsive shoulder. In this case, for low
pressures, the repulsive shoulder can sustain a compact structure with a
lattice parameter related to its range. But when applying enough pressure
the system must collapse to a new compact structure with a lattice parameter
given by the real hard core of the particles.  This kind of models, whether
with a square shoulder, or a linear ramp soft core (which is the one discussed
in this paper) have been studied since long time ago with the picture of core 
collapse in mind.\cite{refe}
Extensions to more general potential were also performed.\cite{refe2} In recent 
papers the problem has been
revisited, and in particular the isostructural transition has been studied
numerically,\cite{refe3} and analytical results have shown that 
in three dimensions, the ground state of a
system with a hard core plus a repulsive shoulder can be one of various
crystalline structures depending on parameters.\cite{refe4}

In this paper I show for the hard core plus linear ramp model in two
dimensions that even the stable zero temperature structures may be very
different from the expected triangular structures. The most stable
configuration may be one of a variety of crystalline structures, and even a
quasicrystal. These structures melt when increasing temperature. The
solid-fluid border in the $P$-$T$ diagram has a zone with negative slope,
which implies a melting with decreasing in volume and in this region the
fluid has an anomalous thermal expansion,\cite{refe5} up to a temperature at which a
density maximum is attained.

The paper is organized as follows. In the next section the model is
introduced and details of the simulation procedure to be used in Section IV
are provided. In Section III the ground state configurations are analyzed.
In section IV, I present detailed results for the $P$-$T$ phase diagram for
a particular value of the parameter $\alpha ,$ which is defined below. In
Section V, possible relevance to real systems are discussed and a summary of
the results is given.

\section{Model and numerical technique}

The model interaction $U\left( r\right) $ between particles that will be
used here consists of a hard core repulsion at a radius $%
r_{0}$ ($U\left( r\right) |_{r<r_{0}}=\infty $), is zero for distances
larger than a value $r_{1}$, and has a soft repulsive part for $r_{0}<r<r_{1}
$ of the form $U\left( r\right) =\varepsilon _{0}\left( r_{1}-r\right)
/\left( r_{1}-r_{0}\right) $ (Figure \ref{potencial}). This interaction
gives a model that is a candidate to have an isostructural transition
between compact configurations of lattice parameter $r_{0}$ and $r_{1}.$ Two
particles interacting through this potential in the presence of an external
force $f$ trying to bring them together, will have a jump in the
interparticle distance from $r_{1}$ to $r_{0}$ when $f$ exceeds the critical
value $\varepsilon _{0}/\left( r_{1}-r_{0}\right) .$ This model is preferred
for numerical simulations instead of the square shoulder model, because it
has much less metastabilities when varying pressure or temperature. If
temperature is measured in units of the energy at contact $\varepsilon _{0}$
(Boltzmann constant is taken to be 1), and distances in units of the hard
core distance $r_{0}$, then $\alpha =r_{1}/r_{0}$ is the only free parameter
of the interaction potential.

\begin{figure}
\narrowtext
\epsfxsize=3.3truein
\vbox{\hskip 0.05truein
\epsffile{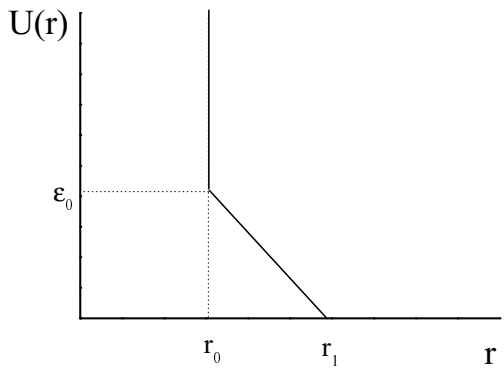}}
\medskip
\caption{The pair potential used throughout the paper. There is a hard core
at distance $r_{0}$ and a soft core (linear ramp) between $r_{0}$ and $r_{1}$%
. Interaction vanishes beyond $r_{1}$.}
\label{potencial}
\end{figure}

Detailed numerical simulations were performed for a two-dimensional system
of 256 particles in the NPT ensemble, using the Montecarlo-Metropolis
technique. A trial movement of a particle consists of a displacement to a
new position chosen randomly inside a cube of a linear size of 1\% of the
mean distance between particles. The new position is accepted with a
Metropolis algorithm, considering the energy change due to the movement.
Once each five Montecarlo swepts through all particles, a trial global
rescaling of all particle coordinates and system size is proposed. The
rescaling is given by a factor chosen randomly within the interval $\pm 0.2\%
$, and is done independently for $x$ and $y$ coordinates, in order 
to allow the system to
accommodate the different crystalline structures that may appear. A maximum
aspect ratio for the system of 1.05 is imposed.
If the trial volume change
does not produce hard core particle overlapping, then it is accepted
according to the Metropolis rule with the value of the energy change $\Delta
E$, given by $\Delta E=P\Delta V-\left( NT/V\right) \Delta V+dE$. Here $N$
is the number of particles, $V$ the volume of the system, $dE$ is the energy
change associated to the change of interparticle distances, and the term $%
-\left( NT/V\right) \Delta V$ $-$which assures the correct limiting equation
of state in the case of an ideal gas (i.e., when $dE=0$)$-$ accounts for the
kinetic energy term that $-$as usual$-$ can be integrated out in the expression
for partition function and thermodynamic potentials. Different runs were
performed at constant pressure starting from random configurations at high
temperature, cooling down to zero temperature and then warming up. Around
10000 Montecarlo steps were used for thermalization at each temperature, and
then 50000 steps were used to calculate thermodynamic quantities, such as
the mean volume $v$ per particle at each temperature, the enthalpy per
particle $h$, the diffraction pattern of the geometrical configurations, and
the diffusion coefficient of the particles $D$.

The diffusion coefficient is calculated in the following way. The distance
traveled by each particle, starting from its initial position, as a function
of time is recorded, and $D$ is taken to be the slope of this function at
long times. From this definition and the kind of simulations performed, it is 
clear that $D$ tends to a constant at high temperatures. The physical 
diffusion cefficient is obtained multiplying by temperature. 
In addition, from the diffraction patterns an orientational
order parameter $B_{m}$ will also be used. It is defined as 
\begin{equation}
B_{m}=\int K\left( k\right) P(k,\theta )\exp (im\theta )d^{2}{\bf k.}
\end{equation}
Here $P\left( k,\theta \right) $ is the intensity of the diffraction pattern
in the ${\bf k}$ plane, in polar coordinates, $m$ is an integer chosen
according the orientational order we are looking for, and $K\left( k\right) $
is a kernel that cuts off the integral at large $k$. The results are
qualitatively insensitive to the form of $K\left( k\right) .$ The expression
used was $K\left( k\right) =\exp (-k^{2})$.

Some comments are in order at this point. The use of the hard core plus
linear ramp potential is motivated $-$as stated before$-$ by numerical reasons.
Both the analytical results of the following section and the numerical 
ones of section III do not change qualitatively if a square shoulder, 
or a parabolic one (with negative second derivative) is used instead of 
the linear ramp. More precise conditions on the potential are discussed 
in Section V.
Results are presented for the two-dimensional case to clarify the
discussions of the structures that will be presented in the next section.
However, the basic properties of the system remain the same in three
dimensions. The nature of the melting transition in two dimensions is a
controversial point in the literature. However, for rather small systems as
the one studied here, indications of discontinuous melting transitions are
clearly observed, but is not obvious whether they remain when the system size
goes to infinity. I will speak throughout the paper of continuous and
discontinuous melting transitions when the simulations indicate each case
for the particular size used in the simulations.

\section{Zero temperature behavior}

\begin{figure}
\narrowtext
\epsfxsize=3.3truein
\vbox{\hskip 0.05truein
\epsffile{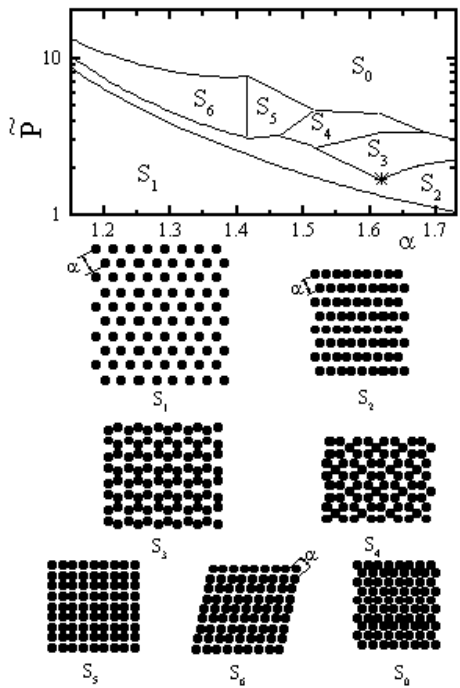}}
\medskip
\caption{Ground state configurations of the system as a function of $%
\widetilde{P}$ ($\equiv Pr_{0}^{2}/\varepsilon _{0}$) and $\alpha .$ At the
point marked by a star, the ground state of the system is a random
quasicrystal. The black dots in the configurations represent the hard core
of the particles.}
\label{alfap}
\end{figure}

Since the first numerical works of Alder and Wainwrigth\cite{aw} in the
sixties, it has been known that the existence of a high enough external
pressure $P$ is sufficient to make a system of otherwise repulsive particles
to freeze. The minimum energy configuration of a system of particles
interacting in two dimensions through a potential of the form $\sim
r^{-\gamma }$ is a triangular lattice for all positive values of $\gamma $.
It is not so widely recognized the great variety of ground states that can
be obtained for more general potentials, even keeping the restriction of a
monotonically decaying potential. We will concentrate in the already
introduced hard core plus linear ramp potential (Fig. \ref{potencial}). The
ground state configuration of a system of particles in two dimensions
interacting through this potential depends on the values of $P$ and $\alpha $%
, and is not necessarily a triangular lattice.

Depending on pressure, nearest particles
tend to be at distance $r_{0}$ or $r_{1}$ from each other. Intermediate
values are not preferred because are not energy minima. The origin of
complex ground state structures in the system is related to the competition
between two terms in the enthalpy $H$ of the system. One is the usual $PV$ term,
which tends to minimize the volume, and the other is the repulsive energy
term, which tends to maximize the interparticle distance. This produces a
sort of frustration, because both terms cannot be minimized at the same
time. The two triangular structures with lattice parameters $r_{0}$ and $%
r_{1}$ (that will be referred to as structures $S_{0}$ and $S_{1}$)
correspond to two ways of reducing the enthalpy by minimizing one term
whereas maximizing the other. These are the best compromises in the case of
very low or very high pressures. However, when both energy terms are
comparable, lower energy intermediate solutions can be found by arranging
the particles with a coordination number (number of neighbors at distance $%
r_{0}$) intermediate between 0 and 6 (which correspond to the structures $%
S_{1}$ and $S_{0}$).

\begin{figure}
\narrowtext
\epsfxsize=3.3truein
\vbox{\hskip 0.05truein
\epsffile{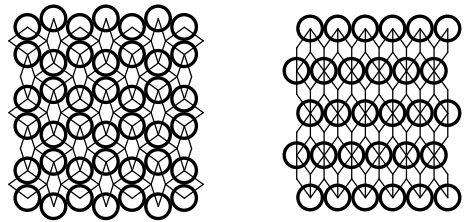}}
\medskip
\caption{Tiling of the structures $S_{3}$ and $S_{2}$ using the
two-dimensional Penrose tiles, for $\widetilde{P}=\widetilde{P}_{\text{qc}}$
and $\alpha =\alpha _{\text{qc}}$. The proportion of thin to fat tiles is
1:3 in $S_{3}$ and 2:1 in $S_{2}$. At $\widetilde{P}_{\text{qc}}$, 
$\alpha _{\text{qc}}$, any possible random tiling of the plane
produces a degenerate ground state.}
\label{ptiles}
\end{figure}

In fact, different crystalline configurations can be proposed, and their
enthalpy calculated in order to find the most stable one as a function of $%
\widetilde{P}$ ($\equiv Pr_{0}^{2}/\varepsilon _{0}$) and $\alpha $. The
result of this analysis is shown in Fig. \ref{alfap}. This figure shows the
results up to a value of $\alpha $ for which the interaction to second
neighbors in the most compact structure ($S_{0}$) is still zero. The
structures in Fig. \ref{alfap} were found by inspection, and they are the
lowest energy configurations found within each region, but other (more
stable) structures may have been missed. Note that some of the structures
have one particle per unit cell (all particles are in translationally
equivalent sites), but in others ($S_{3}$ and $S_{4}$) this number is
greater than one. For some values of $\alpha $ and as a function of $P$,
there are at least three intermediate structure between the triangular ones $%
S_{0}$ and $S_{1}$.

There is one point in the $\widetilde{P}$-$\alpha $ diagram that deserves
further discussion, and this is the one marked by a star in Fig. \ref{alfap}%
. It corresponds to a value of $\alpha \equiv \alpha _{\text{qc}}=1+2\sin
(18^{\text{o}})\cong 1.618$, and $\widetilde{P}\equiv \widetilde{P}_{\text{qc%
}}=1/\sin (36^{\text{o}})\cong 1.7013$. At this point structures $S_{2}$ and 
$S_{3}$ become energetically degenerated, but more importantly, many other
degenerate structures can be constructed. In fact, structure $S_{3},$ and
structure $S_{2}$ for this value of $\alpha $ may be considered as generated
by a tiling of the plane using the two 2D Penrose tiles as indicated in
figure \ref{ptiles}.\cite{libroqc} Particles are located in the far
vertices of the thin tile, and in the nearest ones of the fat tile. At the
point $S_{\text{qc}}$ the enthalpies per particle of the two Penrose tiles
coincide. Any tiling, with the only restriction imposed by the location of
the particles in the above mentioned vertices of the tiles (these are
usually named soft matching rules and the kind of tiling they generate is
known as a random tiling\cite{rt,rt2}) generates a possible ground state of
the system. The proportion of thin to thick tiles used to construct the
ground state is arbitrary (as long as the soft matching rules can be
satisfied). For proportions close to the value corresponding to a perfect
Penrose lattice (nearly 1) the structure obtained is a random quasicrystal.
For pressures lower than $\widetilde{P}_{\text{qc}}$ the structure with a
maximum fraction of thin tiles ($S_{2}$) is preferred, because the thin tile
has lower enthalpy. On the contrary, for pressures higher than $\widetilde{P}%
_{\text{qc}}$ the preferred structure is that with the largest proportion of
fat tiles ($S_{3}$). Quasicrystalline ground states are only stable at the
point $S_{\text{qc}}.$ However, the soft matching rules allow for many ways
of generating them (compared to the more rigid structures $S_{2}$ and $S_{3}$%
), and this implies that at finite temperature the quasicrystalline
structure will be stabilized due to entropic effects, as we will see in the
next section using numerical simulations.

\section{Pressure-Temperature phase diagram}

After having discussed its ground state properties, we focus now on the
behavior of the system at finite temperatures. The main interest is in the
localization of the fluid-solid transition line. I will present now the
numerical results obtained at different values of pressure, in the
particular case $\alpha =1.65$. The system is initialized in a random
configuration at high temperature (well inside the fluid phase) and the
temperature is progressively reduced to zero, and then increased again.

\begin{figure}
\narrowtext
\epsfxsize=3.3truein
\vbox{\hskip 0.05truein
\epsffile{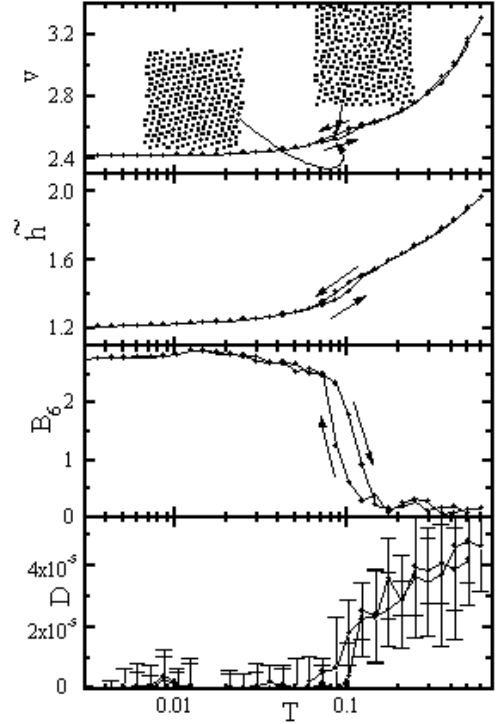}}
\medskip
\caption{Volume $v$, and enthalpy $\widetilde {h}$ ($\equiv E/N+\widetilde{P}v$)
per particle, sixfold orientational order
parameter $B_{6},$ and diffusion coefficient $D$ of the system as a function
of temperature for $\widetilde{P}=0.5$, for a swept decreasing and
increasing temperature ($B_{6}$ and $D$ are given in arbitrary units, $T$ is
in units of $\varepsilon _{0}$). Note the difference in the snapshots of the
system at the same temperature on heating and cooling, within the hysteresis
loop.}
\label{vdet1}
\end{figure}

In the ranges of pressure in which the structures $S_{0}$ and $S_{1}$ are
the most stable zero temperature configurations (according to Fig. \ref
{alfap}), a sharp solidification transition is obtained when reducing
temperature. This can be seen in figures \ref{vdet1},\ref{vdet2}, and \ref
{vdet3} for three different values of pressure within this range. In the
first two the solidification is into the $S_{1}$ structure, and for the
third one into the $S_{0}$ structure. In the three cases the solidification
transition is clearly identifiable by the hysteresis loop in the volume or
the enthalpy of the system, which coincides with the vanishing of the
diffusion coefficient and the appearance of a finite sixfold symmetry of the
diffraction pattern. 

\begin{figure}
\narrowtext
\epsfxsize=3.3truein
\vbox{\hskip 0.05truein
\epsffile{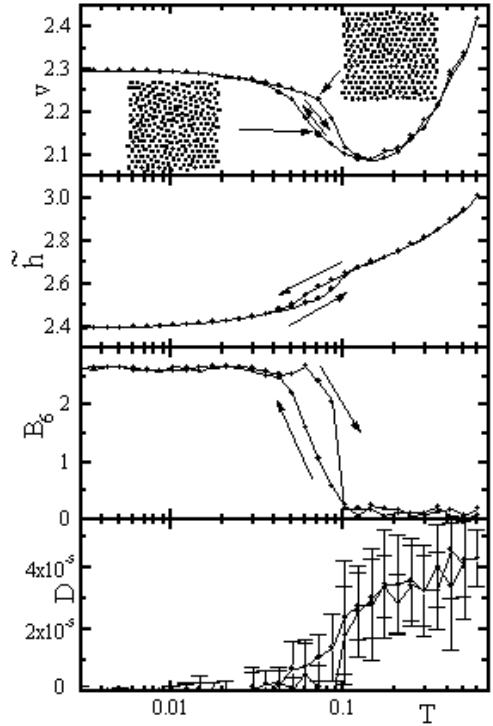}}
\medskip
\caption{Same as Figure \ref{vdet1} for $\widetilde{P}=1.0.$}
\label{vdet2}
\end{figure}

An additional check of the existence of a sharp solid-fluid transition may
be obtained through a long simulation at the equilibrium temperature between
solid and fluid. In this case, the volume of the system should fluctuate
between two clearly different values corresponding to solid and fluid
phases. Results of this simulation are shown for the case $\widetilde{P}=1,$
and $T=0.082$ in Fig. \ref{histo1}. For this simulation the system was
initialized in a fluid equilibrium configuration at the corresponding values
of $\widetilde{P}$ and $T$. After about 5$\times 10^{5}$ Montecarlo steps
the system jumps to the solid phase. After around 2.7$\times 10^{6}$ steps
the systems makes a new short transition to the fluid state.

\begin{figure}
\narrowtext
\epsfxsize=3.3truein
\vbox{\hskip 0.05truein
\epsffile{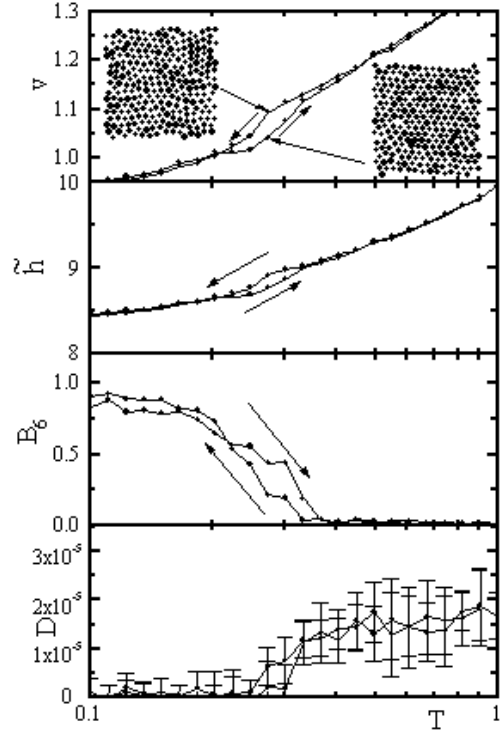}}
\medskip
\caption{Same as Figure \ref{vdet1} for $\widetilde{P}=6.0.$}
\label{vdet3}
\end{figure}

\begin{figure}
\narrowtext
\epsfxsize=3.3truein
\vbox{\hskip 0.05truein
\epsffile{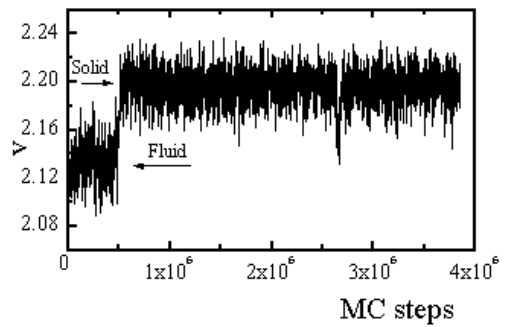}}
\medskip
\caption{Time evolution of the volume per particle $v$ for $\widetilde{P}%
=1.7 $ and $T=0.082,$ close to the fluid-solid transition. After about 5$%
\times 10^{5}$ Montecarlo steps the system jumps to the solid phase. After
around 2.7$\times 10^{6}$ steps the systems makes a new short transition to
the fluid state.}
\label{histo1}
\end{figure}

The most important characteristic to be noted in Fig.\ref{vdet2}, is that
the melting occurs with a reduction in volume for this value of pressure. In
addition, the fluid right after melting has also anomalous thermal expansion
up to some temperature at which a density maximum is attained. These
characteristics imply a negative slope of the solid-fluid coexistence curve,
which is in fact obtained from the simulations as we will see later. The
compressive melting of the system in this region has its origin in the fact
that the usual volume reduction when temperature is reduced is overcome by
the expansion produced when particles diminish their kinetic energy and move
out of the soft core of their neighbors. Illustrating this effect, in Fig. 
\ref{anomfusion} we see snapshots of the system at different temperatures
passing through the liquid-to-solid transition. In the fluid phase there are
particles at distances lower that $r_{1}$ from each other, whereas in the
solid phase the minimum distance between particles is $r_{0}$ (except for
some defects in the structure, which appear mainly because of the
impossibility of accommodate 256 particles in a perfect triangular lattice
within a nearly square cage).

\begin{figure}
\narrowtext
\epsfxsize=3.3truein
\vbox{\hskip 0.05truein
\epsffile{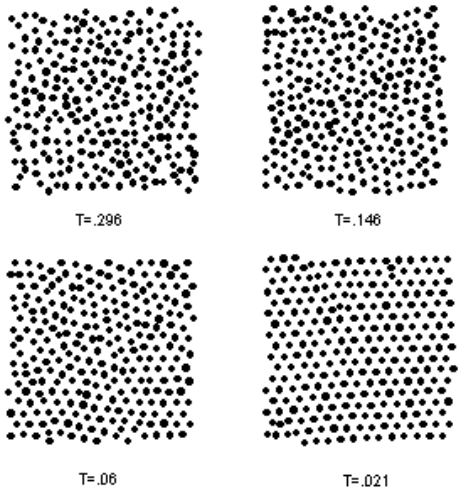}}
\medskip
\caption{Snapshots of the system for $\widetilde{P}=1.7$ when decreasing
temperature, through the fluid-solid transition, illustrating the anomalous
freezing. In the fluid state there are particles at distance lower than $%
r_{1},$ whereas in the solid phase all particles (except a few defects) are
located at distance $r_{1}$ from their neighbors. The temperature at the
second panel corresponds to the maximum density of the system (see Fig. \ref
{vdet2}).}
\label{anomfusion}
\end{figure}

\begin{figure}
\narrowtext
\epsfxsize=3.3truein
\vbox{\hskip 0.05truein
\epsffile{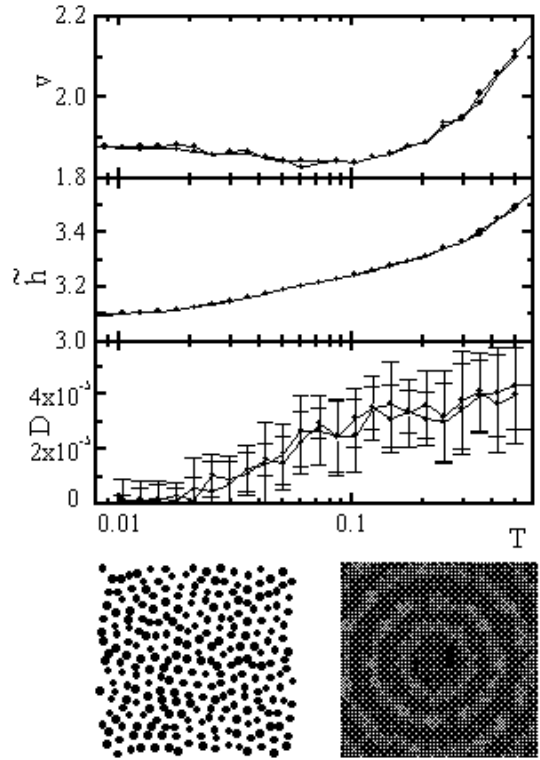}}
\medskip
\caption{Volume per particle $v,$ enthalpy $\widetilde {h}$, 
and diffusion coefficient $D$
of the system as a function of temperature for $\widetilde{P}=1.3$, for a
swept decreasing and increasing temperature. The zero temperature
configuration reached and its diffraction pattern are also shown.}
\label{int1}
\end{figure}

The phases $S_{2},$ $S_{3},$ and $S_{4},$ expected to be the ground states
of the system at intermediate pressures for $\alpha =1.65$, are not
straightforwardly obtained in the simulations. Instead, rather disordered
states are obtained. In figures \ref{int1}, \ref{int2}, and \ref{int3} we
can see the magnitudes $v,$ $\widetilde{h},$ and $D$ for pressures 
$\widetilde{P}=1.3,$
1.7, and 3.8, together with the zero temperature configuration found and the
diffraction pattern of the zero temperature structure. In all the
intermediate pressure range ($1.2\lesssim \widetilde{P}\lesssim 4$) the
enthalpy at zero temperature obtained in the numerical simulation is never
lower than the corresponding to the expected ordered structures of Fig. \ref
{alfap}, as it is shown in Fig. \ref{vdet0}, indicating that probably the
configurations of Fig. \ref{alfap} are really the fundamental states, but
they were not reached in the simulations. The configurations obtained in the
simulations reflect the equilibrium states at some finite but small
temperature, where entropic contributions to the free energy are important,
and they are metastable at zero temperature (note the existence of chains
in Fig. \ref{int1}, and the pentagons and hexagons in Figs. \ref{int2} and
\ref{int3}). Looking at the diffraction
patterns in Figs. \ref{int1}, \ref{int2}, and \ref{int3}, some of them ($%
\widetilde{P}=1.3,$ and $\widetilde{P}=3.8$) show no sign of orientational
order. Others ($\widetilde{P}=1.7$) clearly indicate a tenfold symmetry,
characteristic of a quasicrystal. In the cases where the low temperature
state has no orientational order, the volume or enthalpy of the system do
not show any abrupt solidification transition. In the cases where the low
temperature state has rotational order the volume and enthalpy of the system
show a small hysteretic behavior, suggesting an abrupt solidification
transition. To confirm this fact, long runs were performed at the expected
transition temperature, recording the temporal evolution of volume and
enthalpy. An example of the results for the case $\widetilde{P}=1.7$ is
shown in Fig. \ref{histo2}. The histogram shows a clear bimodal distribution
between two values corresponding precisely to the values of enthalpy
expected from Fig. \ref{int2} at this temperature.

\begin{figure}
\narrowtext
\epsfxsize=3.3truein
\vbox{\hskip 0.05truein
\epsffile{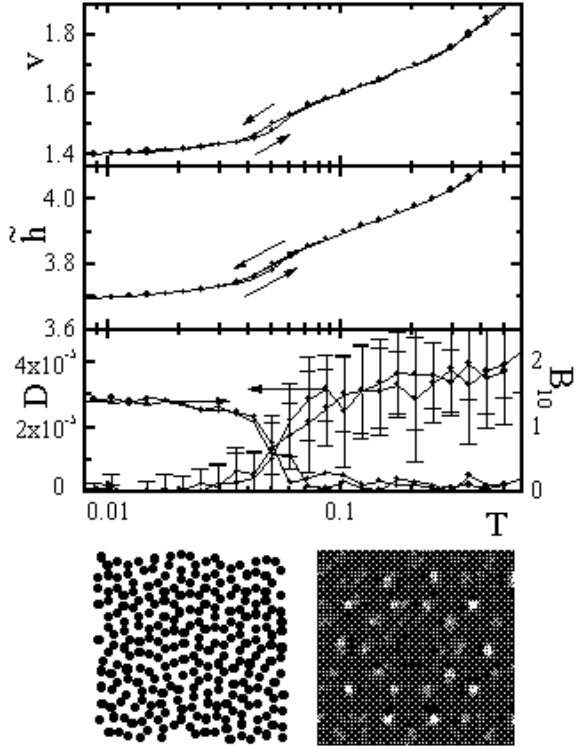}}
\medskip
\caption{Same as Fig. \ref{int1} for $\widetilde{P}=1.7$. In this case the
tenfold orientational order parameter $B_{10}$ is also shown in the last
panel. Note the small histeresis loop in $v$ and $h$.}
\label{int2}
\end{figure}

\begin{figure}
\narrowtext
\epsfxsize=3.3truein
\vbox{\hskip 0.05truein
\epsffile{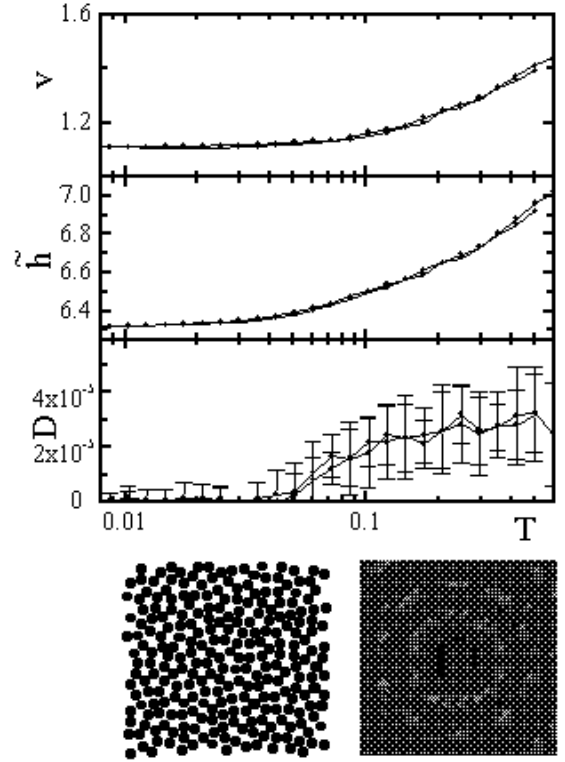}}
\medskip
\caption{Same as Fig. \ref{int1} for $\widetilde{P}=3.8$.}
\label{int3}
\end{figure}

We saw in the previous section that at zero temperature the quasicrystalline
structure is stable only at a particular value of $\widetilde{P}$ and $%
\alpha $. At finite temperatures this structure is stabilized due to
entropic effects, because there are many ways of construct the random
tiling, favoring this structure against the more rigid ones $S_{2}$ and $%
S_{3}$.\cite{tm} A related model of a quasicrystal using two kinds of
particles of different sizes has been studied by Strandburg and coworkers.%
\cite{rt,libroks} In our case, the quasicrystalline state is obtained in a
system of only one kind of particles.

For other values of pressure, the smooth solidification of the system, and
the absence of any obvious order in the low temperature structures obtained,
suggest that the system freezes into a glassy state. However, more detailed
calculations of the diffusion coefficient and other magnitudes in larger
systems are needed to confirm this point.

\begin{figure}
\narrowtext
\epsfxsize=3.3truein
\vbox{\hskip 0.05truein
\epsffile{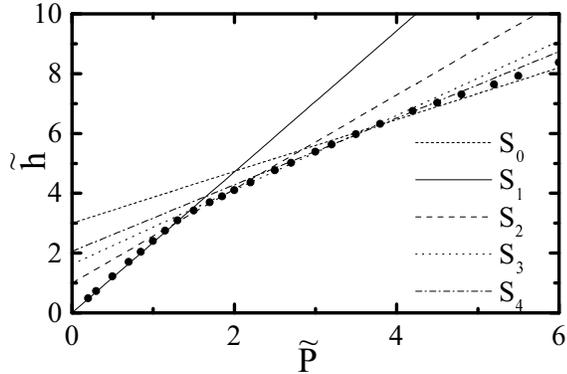}}
\medskip
\caption{Ground state enthalpy per particle as a function of $\widetilde{P}$
from the simulations (black dots) and the analytical expression for the
possible ordered structures. All points lie above at least one of the lines
corresponding to the ordered structures.}
\label{vdet0}
\end{figure}

\begin{figure}
\narrowtext
\epsfxsize=3.3truein
\vbox{\hskip 0.05truein
\epsffile{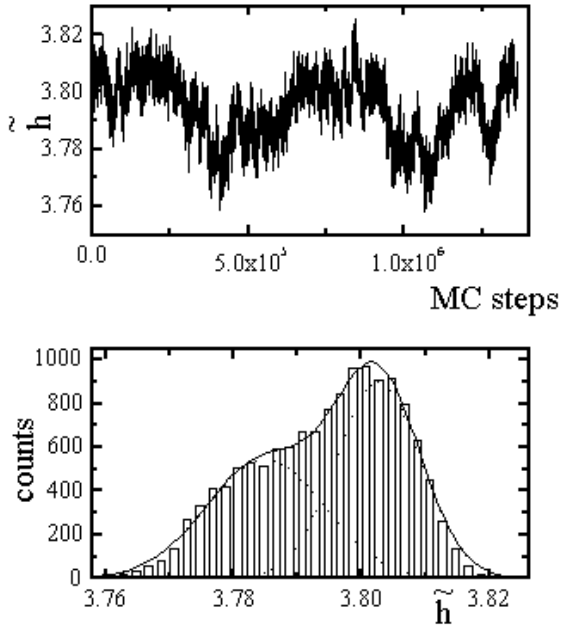}}
\medskip
\caption{Time evolution of the enthalpy per particle for $\widetilde{P}=1.7,$
$T=0.0505,$ close to the fluid-quasicrystal transition. The histogram shows a
clear double peak structure, with mean values compatible with those obtained
from Fig. \ref{int2} for this temperature.}
\label{histo2}
\end{figure}

The numerical results are summarized in the phase diagram of Fig. \ref{pt1}.
The sharp fluid-solid transition in the case of structures $S_{0}$ and $%
S_{1} $ are shown, as well as the corresponding to a quasicrystalline state.
The error bars in these cases are taken as the width of the hysteresis loop
in the enthalpy or the volume of the system. In addition, the approximate
temperature where the system freezes in the other cases is also indicates,
and in this case the error bars indicate the approximate temperature range
in which the diffusion coefficient changes between 10 and 50 \% of its value
at high temperatures.

\begin{figure}
\narrowtext
\epsfxsize=3.3truein
\vbox{\hskip 0.05truein
\epsffile{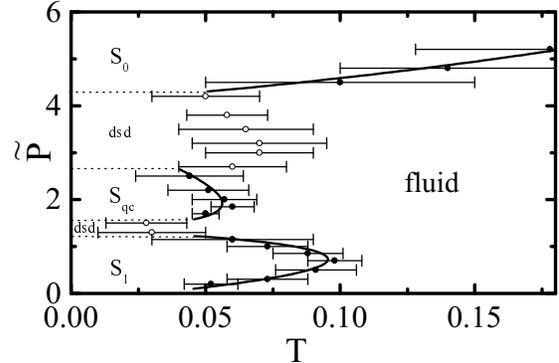}}
\medskip
\caption{Pressure temperature phase diagram from the simulations (dsd stands
for disordered). See text for more details.}
\label{pt1}
\end{figure}

We know from the results of the previous section that this phase diagram
(particularly at intermediate pressures) is metastable at low temperatures.
Since the numerical simulations are not able to reach the fundamental state
in some cases, the determination of the equilibrium phase diagram at all
values of $P$ and $T$ requires the direct comparison of the free energies of
the structures found in the simulations and those known to be more stable in
some cases (structures $S_{2},$ $S_{3}$, and $S_{4},$ for $\alpha =1.65$).
The Gibbs free energy of the system can be obtained from the numerical
simulations through the formula 
\begin{equation}
\frac{G_{2}}{T_{2}}-\frac{G_{1}}{T_{1}}=\int_{1}^{2}\frac{V}{T}dP-\frac{H}{%
T^{2}}dT  \label{dg2}
\end{equation}
where 1 and 2 stand for two set of values $P_{1}$, $T_{1}$ and $P_{2},$ $%
T_{2},$ and the integration is through an arbitrary (reversible) 
path in the $P$-$T$
plane joining points 1 and 2. This determines the free energy of those 
structures obtained in the
numerical simulations up to an overall constant. Also the free
energy of structures $S_{2},$ $S_{3},$ and $S_{4}$ may be determined in this
way, by setting up the configuration of the system at zero temperature in
these structures, and then performing a numerical simulation increasing
temperature. After that, all that remains to be able to compare the free
energies is to fix the additive constants. This was done by introducing in
the model an additional external potential characterized by a strength $W$,
with a periodicity chosen to favor the formation of the required structure.
The reversible path from the ordered structure to that obtained in the
simulation for a given point $P_{0},$ $T_{0}$ consisted of four steps:
increasing $W$ from zero to some large value, increasing $T$ from $T_{0}$ to
a large value, decreasing $W$ down to zero, and decreasing $T$ down to $%
T_{0}.$ The difference in Gibbs free energy $\Delta G$ between the two
structures was calculated through this path by a generalization of the
formula (\ref{dg2}), given by 
\begin{equation}
\Delta \left( G/T\right) |_{P_{0},T_{0}}=\oint \left( \frac{V}{T}dP+\frac{%
E_{W}}{TW}dW-\frac{H+E_{W}}{T^{2}}dT\right) ,  \label{dg}
\end{equation}
where $E_{W}$ is the potential energy of the particles in the artificial
external potential, and $\oint $ indicates the integration along the above
mentioned path. This was done three times with different external potential
to fix all arbitrary constants between free energies of structures $S_{2},$ $%
S_{3},$ $S_{4},$ and the ones obtained in the simulations. After that, the
free energies can be compared and the thermodynamical phase diagram 
constructed.

The complete thermodynamic phase diagram is shown in Fig. \ref{pt3}, where the
stability region of each phase is shown. It is seen that the
quasicrystalline state is in fact thermodynamically stable in a finite range
of $P$ and $T,$ in spite of the value of $\alpha $ ($=1.65$), which is not
the optimum value ($\alpha _{\text{qc}}\cong 1.618$) for the
quasicrystalline structure. Only in the case $\alpha =\alpha _{\text{qc}}$
the quasicrystal is stable down to zero temperature, at the point $\widetilde{P%
}\equiv \widetilde{P}_{\text{qc}}\cong 1.7013$.

\begin{figure}
\narrowtext
\epsfxsize=3.3truein
\vbox{\hskip 0.05truein
\epsffile{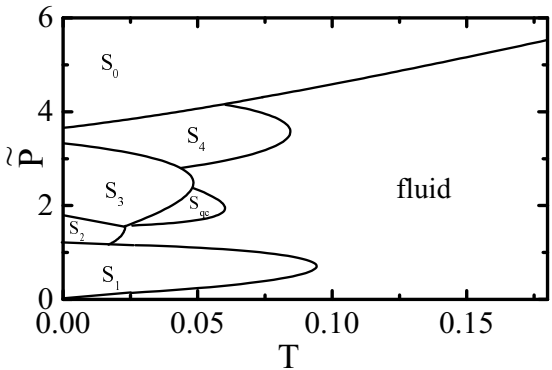}}
\medskip
\caption{Complete pressure-temperature phase diagram for $\alpha =1.65.$ The
errors in the limit of structures $S_2$, $S_3$, and $S_4$ are estimated to be
$\pm 0.01$ in temperature. See
text for discussions.}
\label{pt3}
\end{figure}

\section{Summary and discussion}

In this paper I have discussed the phase behavior of a classical model of
particles interacting through a particularly chosen isotropic potential. The
interaction consists of a hard core plus a linear ramp potential, producing
an effective greater size of the particles at low energies. The ground state
of the system shows different periodic arrangements of the particles,
depending on the values of $P$ and $\alpha $. There is a particular value of 
$P$ and $\alpha $ at which the ground state is a random quasicrystal. These
periodic structures melt when temperature is increased through discontinuous
phase transitions. In addition, at finite temperature, the quasicrystalline
state is stabilized due to entropic effects.

It is usually believed that ``...two-dimensional monoatomic systems
interacting with central forces always form a triangular lattice.''\cite
{malo} Although this is so for power laws and other kinds of interactions,
the ground state configurations of our system show that this is not true in
general. Even considering only crystalline ground states, there may be more
than one atom per unit cell (2 for $S_{3}$ and 5 for $S_{4}$) and 
unequivalent sites within the structure (2 for $S_{4}$). 

Having established the properties of the system for our
model interaction, it is of basic importance to identify
the conditions that an interaction potential must satisfy in order 
to obtain the kind of structures we got in our model. Although it is 
rather difficult to solve the problem in general, something can be 
said about. It is clear that the non-analyticity of the
potential used is not a crucial element, actually, analytic potentials
arbitrary close to the one used here may be easily constructed. 
First of all, let us analyze the necessary 
condition for the local stability of a triangular
structure. Consider particles interacting through a potential $U\left(
r\right) ,$ and distributed in a triangular lattice of lattice parameter $a$. 
At any lattice site the
potential created by all other particles must have a minimum in order for
the structure to be (locally) stable. The potential on that site (taken to
be ${\bf r}=0$) created by a particle at a generic position ${\bf r}$  is, 
up to second order, of the form $U^{\prime \prime }\left( r\right) x^{2}+%
\frac{U^{\prime}\left( r\right) }{r}y^{2}+U^{\prime}\left( r\right) x+%
U\left( r\right)$, where $x$ ($y$) is the coordinate along
(perpendicular to) ${\bf r}$. This potential must be summed up for all
particles, and for lattices with rotational symmetry $C_{3}$ or higher it must
reduce to a isotropic form. Considering the invariance of the trace of 
cuadratic forms
under rotations, the cuadratic part $U_2\left(r\right)$ of the final 
effective potential  can be written as 
\begin{equation}
U_2\left( r\right) =\sum_{i=1}^\infty n_{i}\left[ U^{\prime \prime }\left(
d_i\right) +\frac{U^{\prime }\left( d_i\right) }{d_i}\right] d_i^{2}, 
\label{potencial2}
\end{equation}
where the sum is over all other particles, located at distances $d_{i}$ 
($d_1=a$), and $n_{i}$ is the number of particles at those distances (for
three dimensions the term containing $U^{\prime }$ gets an additional factor
2). The positiveness of this form is the condition for the stability of the
lattice under small displacements of a single particle (global stability is
more difficult to characterize), supposing the lattice parameter is fixed 
from outside. For $U\left( r\right) \sim 1/r^{\gamma }$
all terms of the sum in (\ref{potencial2}) are positive, and the structure is
locally stable. For our
potential, and analyzing a structure with lattice parameter $a$, $r_{1}>a>r_{0}$
, nearest neighbors have $U^{\prime \prime }\left( r\right) =0$ and $%
U^{\prime }\left( r\right) <0,$ and the structure is unstable. 
If the stability condition (\ref{potencial2}) is not satisfied for some
lattice parameter $\widetilde{a}$, then it indicates at least the existence of an 
isostructural transition as a function of pressure between two triangular 
structures with lattice parameters $a_0<\widetilde{a}$ and $a_1>\widetilde{a}$. 
However, 
if first neighbors interaction dominate, it is easy to see that a 
$S_2$ kind of structure is more stable. 
In fact, consider two triangular structures
with lattice parameters $a_0$ and $a_1$ which are degenerate at some pressure
$P$. Their enthalpy per particle must be equal. 
This implies $-$if only first neighbors interaction is important$-$ that
\begin{equation}
Pa_0^2\sqrt{3}/2+3U(a_0)=Pa_1^2\sqrt{3}/2+3U(a_1).
\label{1}
\end{equation}
Now, the enthalpy of a $S_2$ structure with nearest neighbors 
distance $a_0$,
and second neighbors distance $a_1$ is given by
\begin{equation}
Pa_0\sqrt{a_1^2-a_0^2/4}+U(a_0)+2U(a_1),
\label{2}
\end{equation}                          
and it is easy to see that this number is lower than the corresponding to
the triangular structures. We conclude that stable structures other than 
the triangular one will occur if the condition
\begin{equation}
U^{\prime \prime }\left(r\right) +\frac{U^{\prime }\left( r\right) }
{r}<0
\label{3}
\end{equation}                          
is satisfied for some value of $r$. In cases where next-neighbors 
interactions cannot be neglected, a case by case analysis based on 
Eq. (\ref{potencial2}) is needed.

Our model does not contain an attractive part in the potential, and for this
reason it lacks a liquid phase. However a liquid phase can be obtained
within a generalized van der Waals approach,\cite{generalvdv} in which an
attractive interaction is included through an energy term proportional to $-v^{-1}$%
. In addition to the
appearance of a liquid-gas coexistence line, this modification only
renormalizes pressure, and does not affect the structure of the phases, nor
the nature of the anomalies of the phase diagram that were discussed above.

The qualitative features of the phase diagram in the two-dimensional case 
have also been obtained
in simulations with a three-dimensional system. The minimum value of $\alpha 
$ necessary to get the volume anomaly at melting is about $1.2$, both for
two- and three-dimensional systems. 

The results presented here might have importance in another context. 
The anomaly in the fluid-solid
coexistence line, and the negative thermal expansion coefficient in 
this region remind strongly the same effects occurring in water. 
Actually, an interaction potential with a double minimum (which is 
related to the potential considered here) in one dimension has been
analyzed recently, and has been suggested to be the origin of the 
density anomaly in water,\cite{waterprl1} but this proposal has been 
questioned since it seems to give anomalous behaviors only in one dimension.
\cite{waterprl2}
The results presented here show clearly that simple models with anomalous behavior 
may be constructed in two and 
three dimensions. A comparison of the phase diagrams of water
with the corresponding to our system reveals in fact
many coincidences, as the mentioned anomalies and the position of the
different crystalline phases in the $P$-$T$ diagram, which occur near the
pressure where the melting temperature is minimum. In addition,
amorphous structures in ice are well known,\cite{otroice} and the underlying
mechanisms responsible for their formation may be related
to those that originate our disordered structures when cooling down at
intermediate pressures.

Finally, in order to get a better understanding of the dynamic and 
thermodynamic properties of the kind of system we dealt with, 
it would be interesting to find an
experimental realization of an interaction potential satisfying
condition (\ref{potencial2}).
A colloidal system seems to be the natural candidate where to look for 
that realization.

\section{Acknowledgments}

The author thanks J. Simon\'{\i}n and J. Abriata for discussions. This work was
financially supported by Consejo Nacional de Investigaciones Cient\'{\i
}ficas y T\'{e}c\-ni\-cas (CONICET), Argentina.


\begin{references}

\bibitem[\dag]{correo}Electronic address: jagla@cab.cnea.edu.ar

\bibitem{coloides}  W. B. Russel, D. A. Saville, and W. R. Schowalter, {\it %
Colloidal Dispersions} (Cambridge University Press, Cambridge, 1991), 2nd ed.

\bibitem{binary}  W. C. K. Poon and P. N. Pusey, in {\it Observation,
Prediction and Simulation of Phase Transitions in Complex Fluid, }edited by
M. Baus, L. F. Rull, and L. P. Ryckaert (Kluwer Academic Publishers,
Dordrecht, 1994), pp. 3-44; D. H. Napper, {\it Polymeric Stabilization of
Colloidal Dispersions}, (Academin Press, London, 1983), ch. 6.

\bibitem{pocito}  P. Bolhuis and D. Frenkel, Phys. Rev. Lett. {\bf 72}, 2211
(1994); P. Bolhuis, M. Hagen, and D. Frenkel, Phys. Rev. E {\bf 50}, 4880
(1994); C. F. Tejero {\it et al.}, Phys. Rev. Lett. {\bf 73}, 752 (1994); 
{\it ibid.}, Phys. Rev. E {\bf 51}, 558 (1995).

\bibitem{refe}  P. C. Hemmer and G. Stell, Phys. Rev. Lett. {\bf 24},1284
(1970); Stell and P. C Hemmer, J. Chem. Phys. {\bf 56}, 4274 (1972);
J. M. Kincaid, G. Stell, and C. K. Hall, J. Chem. Phys. {\bf 65},
2161 (1976); J. M. Kincaid, G. Stell, and E. Goldmark, J. Chem. Phys. {\bf 65%
}, 2172 (1976); J. M. Kincaid and G. Stell, J. Chem. Phys. {\bf 67}, 420
(1977); J. M. Kincaid and G. Stell, Phys. Letters {\bf 65A}, 131 (1978).

\bibitem{refe2}P. T. Cummings and G. Stell, Mol. Phys. {\bf 43},1267 (1981).

\bibitem{refe3}P. G. Bolhuis and D. Frenkel, J. Phys. Cond. Matter {\bf 9}, 381
(1997).

\bibitem{refe4}C. Rasc\'{o}n {\it et.al.}, J. Chem. Phys. 
{\bf 106}, 6689 (1997).

\bibitem{refe5}In connection with this point see also P. G. Debenedetti, 
V. S. Raghavan, and S. S. Borick, J. Phys. Chem. {\bf 95}, 4540 (1991).

\bibitem{aw}  B. J. Alder and T. E. Wainwright, Phys. Rev. {\bf 127}, 359
(1962).

\bibitem{libroqc}  For a general reference on quasicrystals see C. Janot, 
{\it Quasicrystals} (Clarendon Press, Oxford, 1992).

\bibitem{rt}  C. Henley, J. Phys. A {\bf 21}, 1649 (1988).

\bibitem{rt2}  M. Widom, K. J. Strandburg, and R. H. Swendsen, Phys. Rev.
Lett. {\bf 56}, 706 (1987); K. J. Strandburg, Phys. Rev. B {\bf 60}, 6071
(1989).

\bibitem{tm}  M. Widom, D. P. Peng, and C. L. Henley, Phys. Rev. Lett. {\bf %
63}, 310 (1989).

\bibitem{libroks}  K. J. Strandburg, in {\it Bond Orientational Order in
Condensed Matter Systems}, edited by K. Strandburg (Springer-Verlag, New
York, 1992), ch. 2.

\bibitem{malo}  Reference 10, p. 60.

\bibitem{generalvdv}  P. C. Hemmer and J. L. Lebowitz, in {\it Phase
Transitions and Critical Phenomena, }Vol. 5b, Edited by C. Domb and M. S.
Lebowitz (Academic Press, London, 1976).

\bibitem{waterprl1}  C. H. Cho, S. Sing, and W. Robinson, Phys. Rev. Lett. 
{\bf 76}, 1651 (1996). 

\bibitem{waterprl2} E. Velasco, L. Mederos, and G. Navascu\'{e}s, Phys.
Rev. Lett. {\bf 79}, 179 (1997); C. H. Cho, S. Sing, and W. Robinson, 
Phys. Rev. Lett. {\bf 79}, 180 (1997).

%\bibitem{bismuto}  V. A. Kirilin, V. V. Sychev, A. E. Sheindlin, {\it %
%Engineering Thermodynamics }(p. 174) (Mir Publishers, Moscow, 1976).

\bibitem{otroice}  O. Mishima, L. D. Calvert, and E. Whalley, Nature {\bf 210%
}, 393 (1984); P. Jenniskens and D. F. Blake, Science {\bf 265}, 753 (1994).
\end{references}
\end{document}